# CarGameAR: An Integrated AR Car Game Authoring Interface for Custom-Built Car Programed on Arduino Board

DANG BUI, University of South Florida, USA*

WANWAN LI, University of South Florida, USA*✝

HONG HUANG, University of South Florida, USA

In this paper, we present CarGameAR: An Integrated AR Car Game Authoring Interface for Custom-Built Car Programed on Arduino Board. The car consists of an Arduino board, an H-bridge, and motors. The objective of the project is to create a system that can move a car in different directions using a computer application. The system uses Unity software to create a virtual environment where the user can control the car using keyboard commands. The car's motion is achieved by sending signals from the computer to the Arduino board, which then drives the motors through the H-bridge. The project provides a cost-effective and efficient way to build a car, which can be used for educational purposes, such as teaching programming. Moreover, this project is not limited to the control of the car through keyboard commands in a virtual environment. The system can be adapted to support augmented reality (AR) technology, providing an even more immersive and engaging user experience. By integrating the car with AR, the user can control the car's motion using physical gestures and movements, adding an extra layer of interactivity to the system. This makes the car an ideal platform for game development in AR, allowing the user to create driving games that blend the physical and virtual worlds seamlessly. Additionally, the car's affordability and ease of construction make it an accessible and valuable tool for teaching programming and principles in a fun and interactive way. Overall, this project demonstrates the versatility and potential of the car system, highlighting the various applications and possibilities it offers for both education and entertainment.

## 1 INTRODUCTION

The development of s and automation has become increasingly important in recent years, with the potential to revolutionize many aspects of our lives. In this project, we aim to construct a small car using various components such as the Arduino board, H-bridge, motors, and wheels. Our objective is to create a system that can move the car in different directions using a computer application. To achieve this, we use the Arduino programming language to program the car's movements, allowing it to move forward, backward, left, and right [1]. We utilize a USB connection between our laptop and the Arduino board to send signals from our laptop to the car, instructing it to move according to the input received from the computer. The project provides an efficient and cost-effective way to build a car, which can be used for educational purposes, such as teaching programming and s [2]. Additionally, the car can be used in augmented reality (AR) technology, specifically games, allowing users to control and interact with the car in a virtual environment. Augmented Reality (AR) technology has gained significant attention in recent years due to its potential to enhance user experiences by adding digital elements to the real world [3]. Technology has found a wide range of applications in various fields, including entertainment, education, healthcare, and manufacturing [4]. In particular, AR technology has shown great promise in the field of s, where it has been used to improve the user interface and control of cars. AR technology provides a more intuitive and immersive user experience by overlaying digital information on the physical environment [5]. This technology allows users to interact with digital content in a more natural and seamless way, making it ideal for controlling cars. With AR technology, users can see and control the car in real-time, giving them a better understanding of the car's movements and allowing them to make more informed decisions.

---

* marked as equal contributors
✝ marked as corresponding author

In recent years, there has been a growing interest in combining AR technology with s, as it offers a more advanced and user-friendly way of controlling cars. This integration has led to the development of innovative projects and applications that aim to enhance the capabilities of cars and improve their functionality in various industries [6]. As the field of s continues to grow, it is likely that AR technology will play an increasingly important role in enhancing the user experience and expanding the potential applications of s [7]. The integration of AR technology with cars has enormous potential to revolutionize the field of s. The ability to control and interact with cars through AR provides a more immersive and intuitive experience for users, enabling them to better understand the car's movements and make more informed decisions. However, there are still many obstacles that must be overcome to fully realize the potential of this technology. One of the major challenges is the development of accurate tracking and mapping algorithms that can provide precise and reliable information about the physical environment. Another challenge is the development of user-friendly interfaces that can enable users to control cars with ease [8]. Despite these challenges, there is a growing interest in exploring the potential of this technology, and this project serves as an important foundation for future research and development efforts. By addressing these challenges and building on the foundation laid by this project, it is possible to unlock new possibilities and applications in the field of s.

## 2 RELATED WORK

The paper published by IEEE, "Design of an Arduino-Based Smart Car", presents an extensive analysis of an Arduino-based car that offers two distinct modes of control: manual and automatic. The manual mode allows users to control the car via an Android app connected via Bluetooth. The user can navigate the car using arrows displayed on the app's screen or through voice commands, which the paper mentions as a promising feature. In automatic mode, the car moves in a straight line and uses ultrasonic sensors to scan for obstacles. Upon detecting an obstacle, the car calculates the space available on either side and turns accordingly. Although the IEEE paper presents a promising design, it has the limitation of not being able to use both modes of control simultaneously [9]. In contrast, our project builds on this design and offers additional features through the integration of the Unity program. Our Arduino-based car is controlled through Unity, providing users with an immersive experience of an obstacle avoidance game through augmented reality. With our project, users can experience a more comprehensive and engaging experience by having the added advantage of experiencing the obstacle avoidance game through the AR implementation [10]. Overall, our project extends the design presented by IEEE by offering a more diverse and interactive experience while overcoming the limitations associated with the IEEE project.

The ACM paper titled "Developing an Augmented Reality Racing Game" highlights the development of an AR racing game using an infrastructure designed for the XNA game development platform. The game allows the player to wear a tracked video see-through head-worn display and control the car using a passive tangible controller while other players can manipulate waypoints and obstacles. The authors discuss the game's architecture, software and hardware, user interface, and early demonstrations [11]. The implementation of AR technology in gaming allows for a more immersive and interactive experience for players. In this case, the use of a tangible controller and tracked video display adds to the realism of the game, making it more engaging for the player. However, it is worth noting that implementing such a system can be expensive due to the need for specialized hardware and components, including an AR headset and gameboard [11]. In contrast to the ACM paper, our project aims to provide a more affordable alternative that still offers a high-quality game experience. By utilizing existing hardware and software platforms, we can create a game that is accessible to a wider audience without compromising on the overall experience.

## 3 TECHNICAL APPROACH

**Hardware Construction.** To construct the physical structure of the car, we first designed the chassis using TinkerCAD, a free online 3D design tool. TinkerCAD allowed us to create simple or complex shapes using built-in primitives and imported models from a library [12]. We chose a design that balanced stability and mobility, with enough space to accommodate the necessary components.

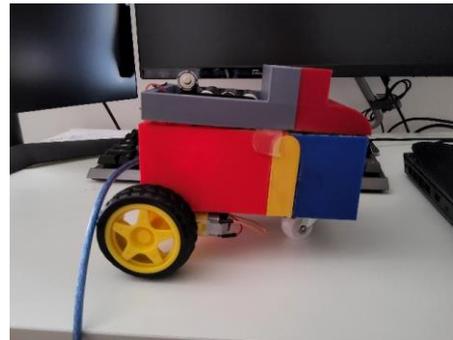

Figure 1: Arduino Car with all components

After designing the chassis, we used a 3D printer to fabricate the parts. The printer used a material called PLA (polylactic acid) to build the layers of the structure according to the design. To ensure that the internal components would fit inside the chassis, we took precise measurements of the available space and the dimensions of the Arduino board, H-bridge, and motors. We made sure to leave enough room for wiring and ventilation, as well as to position the components in a way that would optimize the car's performance [12]. The other required parts to build a simple car can be found in the Arduino starter kit, which includes an Arduino board, an H-bridge, cables, motors, and wheels. These basic components are all that is necessary to construct an affordable car that can be controlled through a computer application. In the car project, the H-bridge plays an important role in controlling the motors. The H-bridge is a circuit that allows us to control the direction of the motor's rotation, as well as the speed of the motor. can be configured in different ways to control the flow of current to the motor.

```
// Executed each frame
void Update()
{
    //This is the code for sending the data
    if (Input.GetKeyDown(KeyCode.A))
    {
        Debug.Log("Sending A");
        serialController.SendSerialMessage("a");
    }

    if (Input.GetKeyDown(KeyCode.W))
    {
        Debug.Log("Sending W");
        serialController.SendSerialMessage("w");
    }

    if (Input.GetKeyDown(KeyCode.S))
    {
        Debug.Log("Sending S");
        serialController.SendSerialMessage("s");
    }

    if (Input.GetKeyDown(KeyCode.D))
    {
        Debug.Log("Sending D");
        serialController.SendSerialMessage("d");
    }

    if (Input.GetKeyDown(KeyCode.H))
    {
        Debug.Log("Sending Stop");
        serialController.SendSerialMessage("h");
    }
```

Figure 2: Snippet of Code in Unity on How It Sends Signal to Arduino

```
void loop() {
  // print out the state of the button:
  if(Serial.available())
  {
    char c = Serial.read();
    if (c)
    {
      if(c == 'w')
      {
        check = "forward";
      }
      else if(c == 'a')
      {
        check = "left";
      }
      else if(c == 's')
      {
        check = "back";
      }
      else if(c == 'd')
      {
        check = "right";
      }
      else if(c== 'h'){
        check = "stop";
      }
      c = NULL;
    }
  }
}
```

Figure 3: Snippet of Code in Arduino IDE on How It Receives Signal from Unity

The H-bridge is connected to the Arduino board, which sends commands to the H-bridge on how to control the motor. It does this by using four switches that when the Arduino board sends a command to the H-bridge, the H-bridge reads it and adjusts the switches accordingly to make the motor spin in the desired direction and at the desired speed. This process is essential for the proper movement of the car, as it allows us to control how the car moves in different directions. Once all the parts were printed and prepared, we assembled the hardware using screws and connectors. We followed the schematic diagrams and instructions provided by the manufacturer of each component to integrate them into the car.

```
if(check == "forward")
{
  digitalWrite(in1, HIGH);
  digitalWrite(in2, LOW);
  digitalWrite(in3, HIGH);
  digitalWrite(in4, LOW);
  digitalWrite(enA, A);
  digitalWrite(enB, A);
  delay(500);
  analogWrite(enB, 100);
  analogWrite(enA, 100);
  Serial.println("Straight");
}
else if (check == "back")
{
  digitalWrite(in1, LOW);
  digitalWrite(in2, HIGH);
  digitalWrite(in3, LOW);
  digitalWrite(in4, HIGH);
  analogWrite(enA, A);
  analogWrite(enB, A);
  delay(500);
  analogWrite(enB, 100);
  analogWrite(enA, 100);
  Serial.println("Back");
}
```

Figure 4: Snippet of Code in Arduino IDE on How It Executes the Code.

**Software Construction.** The software used in the construction of the car includes Arduino IDE and Unity. To enable bidirectional communication between the Arduino board and Unity through COM port, the Ardity package was downloaded and set up [13]. In Unity, the user input is recorded and converted into a message that is sent to the Arduino board for further processing. For instance, 'w' is interpreted as the car moving forward, 'a' as the car turning left, 's' as the car moving backward, and 'd' as the car turning right. Upon receiving the signal from Unity, the Arduino board processes the instruction and sends the command to crucial components like the motor, causing it to spin in a particular direction as per the input message [14]. Upon receiving the signal from the Unity program, the Arduino IDE will interpret that information and save the instruction to the variable check.

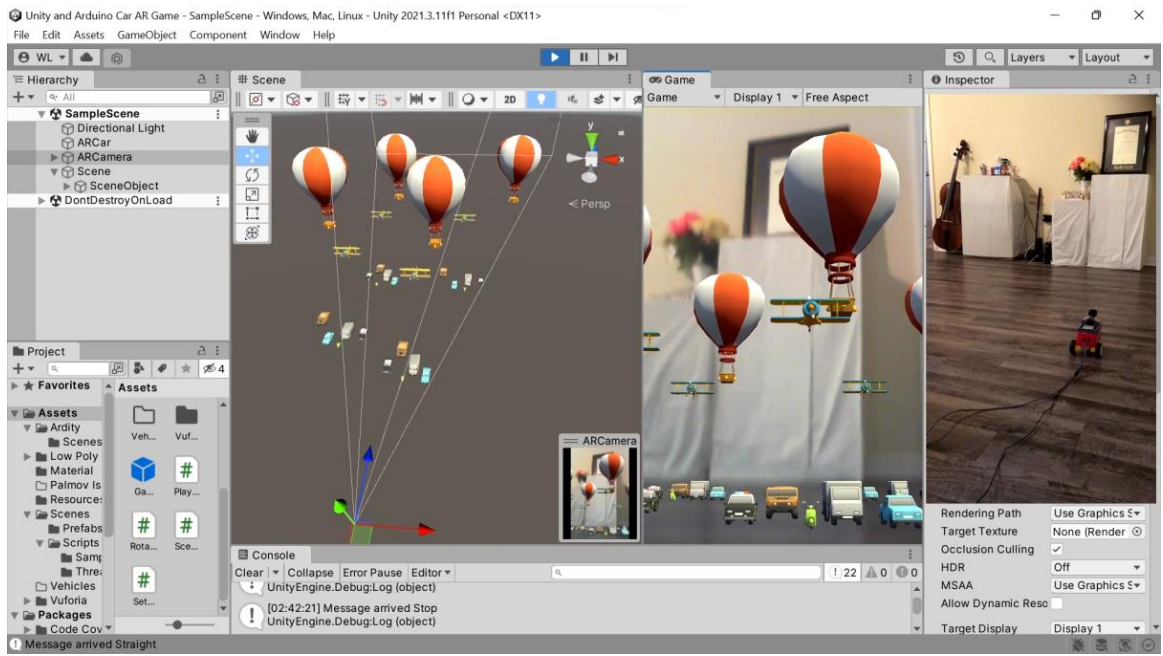

Figure 5: Screenshot of experiment results that tests AR Arduino car on the game authoring interface.

The Arduino board also sends back a signal to Unity to confirm whether it has received the input signal and has successfully carried out the instruction. This bidirectional communication between the two software enables the car to move and perform specific tasks as instructed by the user [15].

## 4 EXPERIMENT RESULT

We have conducted the experiment with the Arduino car on both a smooth surface as well as a rough surface. Figure 5 shows the screenshot of the experiment results that test the programmed Arduino car on the AR game authoring interface. When the W key was pressed, the car was observed to move forward at a slow to medium speed. Subsequently, when the A key was inputted into the Unity program, the right motors were noted to spin faster, resulting in a left turn by the Arduino car. Similarly, pressing the D key caused the left motors to spin faster, resulting in a right turn. Finally, when the H key was pressed, the car was observed to come to a stop. The results showed that the car performed better on a flat surface where the contact area between the car and the surface is even. This allowed each of the motors and wheel to perform normally, resulting in smoother and more accurate movement of the car. However, on a rough surface, the car tended to move crookedly, not in the desired direction. In the condition of a rough surface, there would be more friction between the wheel and the surface, which can cause the motor to spin slower, resulting in a decrease in speed and power of the car. This is due to the fact that the wheels have to work harder to maintain contact with the uneven surface, which can cause the car to move crookedly, not in the direction that we want.

Additionally, when the weight of the car is not distributed evenly between all the wheels, it can affect the balance and stability of the car, making it more difficult to control. As a result, the experiment suggests that the performance of the car is highly dependent on the surface on which it operates and that a flat and even surface is necessary for optimal performance. The main connection between the car and the computer was through a long USB cable, occasional delays were observed in the transfer of signals from the computer to the car. During the experiment, we encountered some unexpected challenges with the original design of the car. Despite adjusting the motor speed, the car's movement r emained unpredictable, with no consistent turning direction. It became apparent that the car was not in a balanced state, which caused the caster wheel to spin randomly, throwing off the car's trajectory. As a quick fix, we removed the caster wheel and replaced it with two additional wheels, effectively transforming the car into a 4-wheel drive. This solution initially appeared to work well, with the car moving smoothly and accurately in the intended direction. However, we soon discovered that the battery consumption was high, resulting in the car's performance deteriorating over time. The car needed new AA batteries every time it was used to ensure the best performance, making the 4-wheel drive solution impractical for long-term use. As a result, we explored alternative solutions to address the balance issue and improve the car's overall performance.

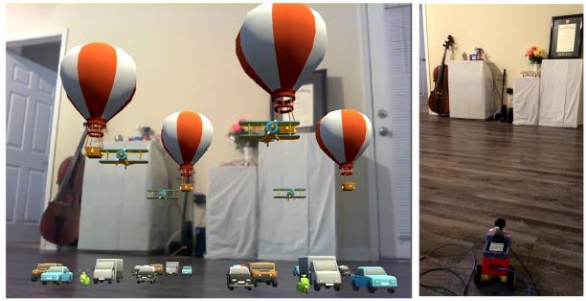

Figure 6: Game view of AR Arduino car game

After the battery issue arose with the 4-wheel drive solution, we delved deeper into the car's design to find a more sustainable solution. We suspected that the car's weight distribution could be a factor affecting its balance and movement. To test this hypothesis, we tried adding additional weights to the front of the car to improve its balance. While this helped to some extent, it did not entirely solve the problem of the car's unpredictable movement. Next, we examined the wiring connections between the H-bridge and motor and found that they were loose, causing signal interference and further contributing to the car's instability. We used a glue gun to reinforce these connections, and this resulted in a noticeable improvement in the car's performance. After exploring alternative solutions to address the balance issue and improve the car's overall performance, we decided to try using a rechargeable lithium-ion battery instead of disposable AA batteries. While this did save some battery power, we quickly realized that it did not provide enough power for the car to perform left or right turns, which require two motors on one side to spin much faster than the other. Realizing that this was not a viable solution, we decided to revert to the original design with one caster wheel in the front and two wheels at the back. Upon testing the car with this setup, we discovered that the caster wheel could spin in any direction, causing the car to move in that direction. Since the car was designed to perform tasks such as going straight, left, right, and back, we needed to prevent the caster wheel from spinning randomly. To address this issue, we decided to glue the caster wheel to the chassis so that it could only rotate forward and backward. Not only did this change save battery power by eliminating the need for two extra motors, but it also allowed the car to move in the desired directions of straight, left, right, and back. Figure 6 shows the game view of the AR Arduino car game. Overall, the Arduino car was observed to perform satisfactorily in accordance with the input signals from the computer. For more details on the experiment result, please refer to this video: https://youtu.be/oG91jNgPDuQ

## 5  CONCLUSION

The development of technologies such as augmented reality (AR) has led to the creation of new and exciting experiences in various fields. In the field of gaming, AR has allowed for the creation of interactive and immersive experiences that were previously impossible. In this paper, we present CarGameAR: An Integrated AR Car Game Authoring Interface for Custom-Built Car Programed on Arduino Board.  This paper shows the implementation details of CarGameAR and opens a new discussion about its potential applications, and its impact on the gaming industry. CarGameAR allows users to create their own custom-built cars and program them with Arduino boards. The programming aspect of CarGameAR allows users to create custom behavior and movements for their cars. The AR-based game creation feature of CarGameAR allows users to create immersive games that utilize their custom-built cars. The game creation process is simplified through the use of a drag-and-drop interface in Unity3D Editor, which allows users to easily add game elements such as obstacles, trees, stones, etc. The AR technology allows for the game to be displayed in a real-world environment, adding a new layer of immersion to the gaming experience.

**Future Work.** As future work, CarGameAR has the potential to be used in various applications, including educational settings. The tool can be used to teach students about programming, engineering, and physics concepts in a fun and interactive way. It can also be used as a tool for team-building activities, allowing teams to work together to create custom cars and compete in AR-based games. CarGameAR can also be used in marketing and advertising campaigns. Companies can create custom-branded cars and use them in AR-based games to promote their brand. This can be a fun and engaging way to increase brand awareness and reach a younger demographic. On the other hand, CarGameAR has the potential to revolutionize the way we play car games. CarGameAR allows for a high degree of customization, allowing players to create their own unique Ardruino cars and behaviors. In all, CarGameAR is a unique and innovative interface that allows for the creation of custom-built cars programmed on Arduino boards, as well as the creation of AR-based games utilizing these cars. Its potential applications and impact on the gaming industry make it an exciting development in the field of AR and game development.